1

# The Minerageny of Two Groups of Zircons from Plagioclase-Amphibolite of Mayuan Group in Northern Fujian

Xuezhao Bao[a1] and Gan Xiaochun[b]

[a]Department of Earth Sciences, the University of Western Ontario, 34-534, Platt's Lane, London, Canada, N6G 3A8
[b]Tianjin Institute of Geology and Mineral Resources, Tianjin, China 300170

Zircons can crystallize in a wide range of physical and chemical conditions. At the same time, they have very high stability and durability. Therefore zircons can grow and survive in a variety of geological processes. In addition, the diffusivity of chemical compositions in their crystals is very low. Consequently, we can trace back the evolution history of the planetary materials containing zircon with zircon U-Th-Pb geochronology and geochemistry studies. However, this depends on our ability to decipher its genesis, namely magmatic or metamorphic origins. In this paper, magmatic and metamorphic zircons were found from plagioclase-amphibolite samples. Their geneses have been determined by zircon morphology, chemical composition zonations and geological field settings combined with their zircon U-Th-Pb ages. We have found obvious differences in micro-scale Raman spectra between these magmatic and metamorphic zircons. The magmatic zircons exhibit a high sloping background in their Raman spectra, but the metamorphic zircons exhibit a low horizontal background in their Raman spectra, which suggest that the magmatic zircons may contain a much higher concentration of fluorescent impurities than the metamorphic zircons. Moreover, reverse variation trends in Raman spectrum peak intensities from core to rim of a crystal between the magmatic and metamorphic zircons have been found. We think that this can be attributed to their reverse chemical composition zonations. These differences can be used to distinguish magmatic and metamorphic zircons.

Key words: magmatic and metamorphic zircons; electron microprobe; Micro-Raman spectrum; plagioclase amphibolite, zircon genesis identification methods.

## 1. THE GEOLOGICAL OCCURRENCE, CRYSTAL SHAPES AND U-Pb AGE OF ZIRCONS

The zircons were extracted/ separated from plagioclase amphibolite samples taken from an area close to Lvkou village, 20km west of Jianyang County in northern Fujian, China. It belongs to the lower member of the Dajinshan formation in the Mayuan Groups. These groups of rocks consist of mainly biotite - plagioclase metamorphic rock, with a small amount of plagioclase amphibolite and marble. The index metamorphic minerals include sillimanite, kyanite, garnet and diopside, which indicate that these rocks have experienced an amphibolite facies metamorphism. The geological field setting has suggested that their parent rocks have formed in the early Proterozoic era [1]. The major minerals in this plagioclase-amphibolite are amphibole and plagioclase (andesine or oligoclase), with a small amount of diopside. Two groups of zircons have been identified:

---

[1] Corresponding author. E-mail address: xuezhaobao@hotmail.com (Xuezhao Bao)

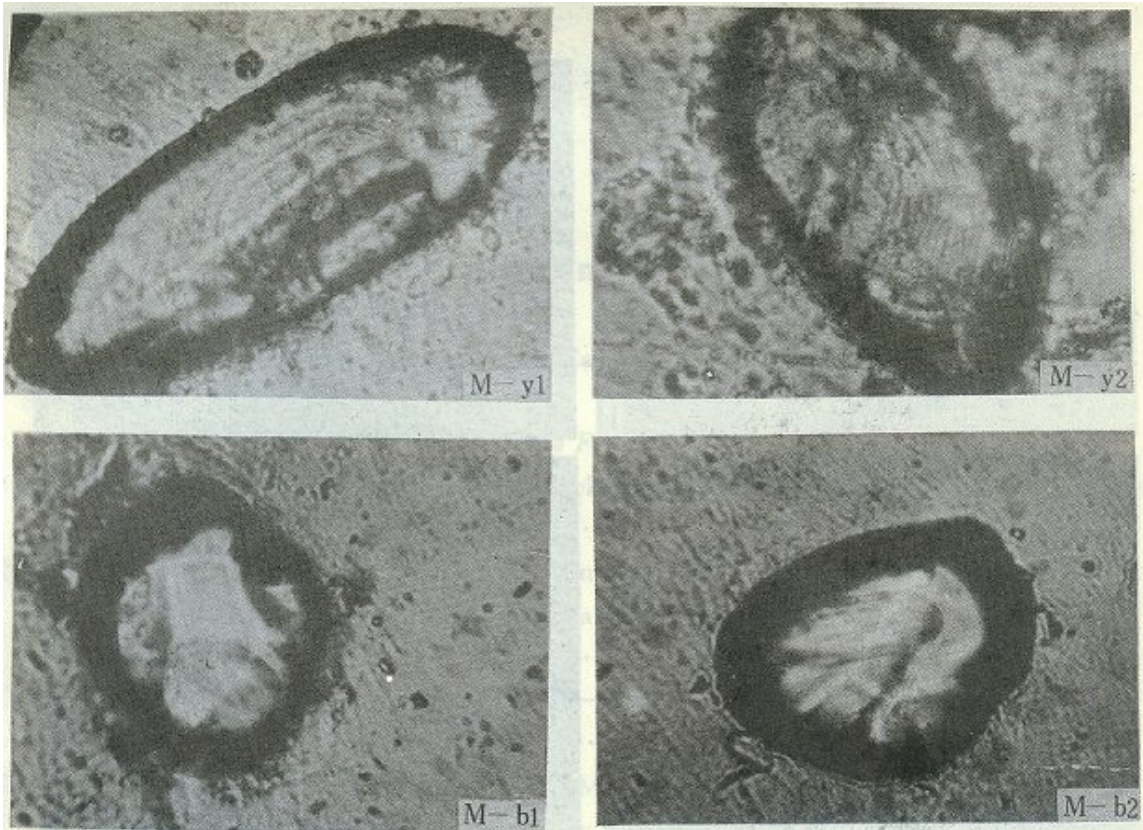

**Fig. 1. The microscopic photos of zircons**
**M-y1 and M-y2 with with euhedral growth zonings. Magnification X400.**
**M-b1 and M-b2 with rounded shape and core. Magnification X400.**

Zircons M-y: have rounded, long columnar crystal shapes, brown-red color, sizes of 0.3-0.5 mm. They exhibit euhedral growth zonings as shown in pictures M-y1 and M-y2 in fig.1, which is one of the features of magmatic zircons [2] Their U-Th-Pb single zircon age is 2300 million years (Ma)± (all of the U-Th-Pb single zircon ages in this paper were analyzed in the Geochronology and Isotope Geochemistry Lab, Tianjin Institute of Geology and Mineral Resources).

Zircons M-b: exhibit equant rounded crystal shapes, are colorless, transparent, and have sizes of 0.1-0.25 mm, and rounded nuclei in their crystals as shown in pictures M-b1 and M-b2 in fig.1, which is consistent with the features of metamorphic or metamorphic re-crystallization zircons [3]. Their U-Pb single zircon age is 400Ma±.

Since the parent rock formed in the early Proterozoic era [1], the first group of zircons crystallized from magma at 2300 Ma±, and the second group of zircons crystallized in a metamorphic event that occurred at 400 Ma±.

## 2. THE STRUCTURE AND COMPOSITION ZONATION FEATURES

### 2.1 Laboratory works

#### 2.1.1 Electron microprobe

These zircons were mounted in epoxy resin, and then grinded and polished for electron microprobe (EMP) analysis. A JEOL JXA-8600 EMP was used to analyze the chemical composition through the zircon crystals. The analysis results are listed in Table 1.

2.1.2 Raman spectrum micro analysis

　　After EMP analysis, these samples were re-polished. A T6400 Raman spectrometer made by France's J.Y. Company was used to analyze these zircons. Analysis conditions are: micro-Raman x10 camera lens; CCD detector; accumulated analysis time: 150 seconds/analysis point; optical slit: 200/250; the wavelength of laser beam: 514.5 nm; the power of laser beam: 200 nw; the spot size of laser beam can be adjusted to 2 μm in diameter. Analyzer: Luo Wei, Raman spectroscopy lab, Jilin University.

**Table 1. Chemical composition from core to rim (1→4) of zircon crystals**

| sample # | spot # | SiO2 | UO2 | ThO2 | ZrO2 | HfO2 | FeO | P2O5 | Y2O3 | Σ | ZrO2/HfO2 |
|---|---|---|---|---|---|---|---|---|---|---|---|
| M-y1 | 1 | 33.04 | 0.15 | 0.00 | 65.23 | 1.21 | 0.25 | 0.07 | 0.00 | 99.95 | 53.91 |
| | 2 | 33.31 | 0.00 | 0.00 | 66.35 | 1.33 | 0.14 | 0.07 | 0.00 | 101.10 | 49.89 |
| | 3 | 32.95 | 0.00 | 0.00 | 67.07 | 1.36 | 0.16 | 0.03 | 0.07 | 101.64 | 49.32 |
| | 4 | 33.21 | 0.08 | 0.00 | 66.00 | 2.19 | 0.30 | 0.00 | 0.00 | 101.78 | 30.14 |
| M-y2 | 1 | 33.68 | 0.07 | 0.09 | 67.69 | 1.17 | 0.06 | 0.10 | 0.15 | 103.01 | 57.85 |
| | 2 | 33.54 | 0.06 | 0.00 | 65.17 | 1.78 | 0.00 | 0.00 | 0.00 | 100.55 | 36.61 |
| | 3 | 32.68 | 0.06 | 0.07 | 65.33 | 1.79 | 0.07 | 0.00 | 0.00 | 100.00 | 36.50 |
| | 4 | 33.75 | 0.00 | 0.05 | 64.66 | 1.54 | 0.00 | 0.00 | 0.00 | 100.00 | 41.99 |
| M-b1 | 1 | 32.66 | 0.00 | 0.00 | 65.82 | 1.47 | 0.00 | 0.05 | 0.00 | 100.00 | 44.78 |
| | 2 | 32.34 | 0.00 | 0.00 | 66.42 | 1.21 | 0.00 | 0.03 | 0.00 | 100.00 | 54.89 |
| | 3 | 33.05 | 0.14 | 0.00 | 67.03 | 1.10 | 0.00 | 0.03 | 0.05 | 101.40 | 60.94 |
| | 4 | 32.53 | 0.00 | 0.00 | 65.97 | 1.24 | 0.00 | 0.05 | 0.00 | 99.79 | 53.20 |
| M-b2 | 1 | 31.75 | 0.08 | 0.00 | 66.70 | 1.45 | 0.00 | 0.02 | 0.00 | 100.00 | 46.00 |
| | 2 | 33.15 | 0.00 | 0.00 | 67.42 | 1.17 | 0.00 | 0.00 | 0.00 | 101.74 | 57.62 |
| | 3 | 32.41 | 0.00 | 0.00 | 67.82 | 1.02 | 0.00 | 0.03 | 0.00 | 101.28 | 66.49 |
| | 4 | 31.59 | 0.00 | 0.00 | 65.98 | 1.45 | 0.00 | 0.00 | 0.05 | 100.17 | 46.19 |

**Analysis conditions: voltage 15 KV; current 2 x 10$^{-8}$ A.**

**　　Analyzer: Fang Guan, Electron microprobe lab, Beijing Institute of Geology, the ministry of Nuclear power industry of China.**

**　　The main standards that have been used are: zircon for Zr, 100% metal Hf for Hf, $UO_2$ for U, $ThO_2$ for Th, $KVPO_3$ for P, $Y_3Al_5O_{12}$ for Y and $FeS_2$ for Fe.**

## 2.2 Results and discussion

2.2.1 Chemical composition zonation features

　　Zircons M-y exhibit a trend of increasing $HfO_2$ concentration and a trend of decreasing $ZrO_2/HfO_2$ ratio from the centre to the rim of a crystal (the abnormality in the rim of M-y2 zircon may be caused by later metamorphism [4]), which is consistent with the features of magmatic zircons [4-5], supporting a magmatic origin.

　　M-b group zircons exhibit a trend of decreasing $HfO_2$ concentration and a trend of increasing $ZrO_2/HfO_2$ ratio from the centre to the rim of a crystal (the abnormality in their rims may be caused by later retro-metamorphism[4]), which is consistent with the features of metamorphic zircons [4-5]. These characteristics support their metamorphic origin.

2.2.2 Micro Raman spectrum features

　　There are significant differences in Raman spectra between the magmatic

**Table 2. Observed vibrational Raman frequencies (cm$^{-1}$) from core to rim of magmatic and metamorphic zircons**

| Genesis | Sample # | assignment / Spot # | Eg(III)* | B$_2$g(II) | Eg(I) | Eg(v4) | A$_1$g(v2) | A$_1$g(v1) | B$_1$g(v3) | |
|---|---|---|---|---|---|---|---|---|---|---|
| Magmatic | M-y1 | core 1 | 202.09 | | 224.23 | 355.43 | 437.64 | 972.48 | 1005.30 | 1054.1 | 1087.70 |
| | | 2 | 201.58 | | 224.23 | 355.43 | 438.25 | 972.58 | 1006.10 | 1054.50 | |
| | | 3 | 201.68 | | 224.23 | 355.02 | 438.25 | 972.48 | 1005.70 | 1054.10 | |
| | | rim 4 | 201.68 | 219.76 | 224.23 | 355.43 | 438.25 | 972.58 | 1006.10 | | |
| | M-y2 | core 1 | 199.89 | | 221.77 | 353.38 | 435.79 | 970.84 | 1004.60 | 1052.80 | |
| | | 2 | 200.04 | | 221.19 | 352.56 | 435.79 | 970.84 | 1004.00 | 1052.40 | |
| | | 3 | 199.63 | | 221.77 | 353.38 | 436.20 | 970.84 | 1004.50 | 1052.80 | |
| | | 4 | 199.23 | | 221.77 | 353.38 | 436.20 | 971.25 | 1004.50 | 1053.20 | |
| | | rim 4 | 200.04 | | 221.77 | 353.98 | 436.20 | 970.84 | 1004.50 | | |
| | 9303 | core 1 | 202.09 | | 225.05 | 357.46 | 437.64 | 974.12 | 1007.30 | 1055.30 | |
| | | 2 | 202.09 | | 224.64 | 355.64 | 438.25 | 973.30 | 1008.50 | 1054.50 | |
| | | 3 | 201.68 | 213.16 | 224.23 | | 437.64 | 972.89 | 1007.30 | 1054.50 | |
| | | rim 4 | 201.68 | 213.16 | 223.82 | | 437.64 | 973.30 | 1005.90 | 1054.50 | |
| Metamorphic | M-b1 | core 1 | 200.04 | | 223.00 | 355.02 | 438.25 | 973.30 | 1006.90 | 1053.20 | |
| | | 2 | 199.22 | 211.93 | 222.18 | 354.20 | 437.02 | 972.07 | 1005.70 | | |
| | | 3 | 200.45 | 213.16 | 223.41 | 355.43 | 438.25 | 973.71 | 1007.30 | 1053.70 | |
| | | rim 4 | 200.86 | 213.57 | 223.82 | 355.43 | 438.66 | 973.71 | 1007.30 | 1054.90 | |
| | M-b1 | core 1 | 198.40 | 210.29 | 221.36 | 353.93 | 436.20 | 971.25 | 1004.50 | 1052.40 | 1087.30 |
| | | 2 | 198.81 | 210.70 | 221.77 | 353.78 | 436.20 | 971.66 | 1004.90 | 1052.40 | |
| | | 3 | 199.22 | 211.52 | 222.18 | 354.20 | 436.61 | 971.66 | 1005.30 | 1053.20 | |
| | | 4 | 198.81 | 211.52 | 222.18 | 353.79 | 436.20 | 971.66 | 1004.90 | 1052.80 | |
| | | rim 5 | 199.63 | 211.52 | 222.59 | 354.61 | 436.61 | 972.07 | 1005.70 | 1054.10 | |
| | 87013 | core 1 | 202.02 | 213.96 | 225.06 | 356.25 | 439.48 | 974.59 | 1007.70 | 1055.90 | 1090.10 |
| | | 2 | 201.68 | 213.57 | 224.64 | 356.25 | 439.48 | 974.12 | 1007.30 | 1050.90 | |
| | | 3 | 202.09 | 213.98 | 224.64 | 356.25 | 439.48 | 974.12 | 1007.70 | 1054.90 | |
| | | 4 | 201.68 | 213.57 | 224.64 | 356.25 | 438.07 | 974.59 | 1007.70 | 1055.90 | |
| | | 5 | 202.50 | 214.39 | 225.05 | 356.66 | 439.48 | 974.94 | 1008.10 | 1055.70 | |
| | | rim 6 | 202.09 | 214.98 | 224.64 | 356.25 | 439.07 | 974.59 | 1008.10 | 1055.90 | |

*Raman spectrum peak names from ref. 6.

and metamorphic zircons. The observed vibrational frequencies of Raman spectra from core to rim of zircons are listed in table 2. The typical Raman spectra of zircons M-y were shown in fig.2. For comparison, the Raman spectra of a typical magmatic zircon (No. 9303) from the pegmatite samples located in Xiao Qingling [4] is shown in fig.3. The typical Raman spectra of zircons M-b is shown in fig.4. Zircon 87013, coming from Danzhu granite, has been confirmed as a metamorphic zircon in previous studies [3,4]. Its Raman spectra are shown in fig.5 for comparison. The Raman spectrum peaks of zircons have been assigned in ref 6. Several noticeable points were found:

(1) The Raman spectra of the magmatic zircons have high sloping backgrounds. Their multi-crystal powder samples also exhibit the same Raman spectrum characteristics. On the other hand, the metamorphic zircons crystallized from rocks that have undergone medium to low grade metamorphism exhibit low horizontal backgrounds in their Raman spectra. Similarly, their powder samples also exhibit the same Raman spectrum characters.

Possible mechanisms: Raman spectra can be heavily influenced by fluorescent impurities, even if their concentrations are below the detection of an electron microprobe [7]. These fluorescent impurities usually are transition metals and rare earth elements, such as Eu, La, Ca, Y etc. [8]. These impurities can enter the zircon crystals, and their concentrations are origin dependent (their concentrations can differ due to their origins) [9]. Therefore, a high background level of the Raman spectra of the magmatic zircons may originate from a high concentration of fluorescent impurities, but a low background of the Raman spectra of metamorphic zircons may originate from a low or zero concentration of fluorescent impurities. For magmatic zircons, a relatively high concentration of fluorescent impurities may come from the following two factors: 1) magma is a high temperature melt with complex chemical compositions induced by its high temperatures, therefore it may contain a relatively high concentration of fluorescent impurities. In addition, it also can absorb fluorescent impurities from its surrounding rocks. Therefore, zircons crystallized in magma have more chances to absorb fluorescent impurities, such as REE, into their crystals. 2) High temperatures are favorable for elements to enter a crystal by solid solution mixing. Consequently, the high temperature of magma allows the zircon crystals to absorb more fluorescent impurities. When compared with magmatic zircons, the metamorphic zircons crystallized at a relatively lower temperature. For instance, the crystallization temperature of zircons M-b is 600~700 $^{\circ}$C [1]. The parent rock of zircon 87013 only experienced a medium grade of metamorphism. Therefore, its crystallization temperature should not be so high (< 700 °C). Under these conditions, it may be difficult for the fluorescent impurities to enter the lattice of zircons. In addition, the chemical composition of medium-low metamorphic fluid that zircons crystallized from is not as complex as that of magma because of its low temperature. This may be another cause leading them to have low levels of fluorescent impurities (notes: higher background levels have been found in the Raman spectra of zircons from high temperature metamorphic rock, granulites. One of these instances is the zircon G2126 [4,5] from a granulite in the upper Jiling group located in Xinghe, inner Monglia. It has a relatively higher background level in its Raman spectra, which may be produced by a higher concentration of fluorescent impurities induced by a higher metamorphic temperature).

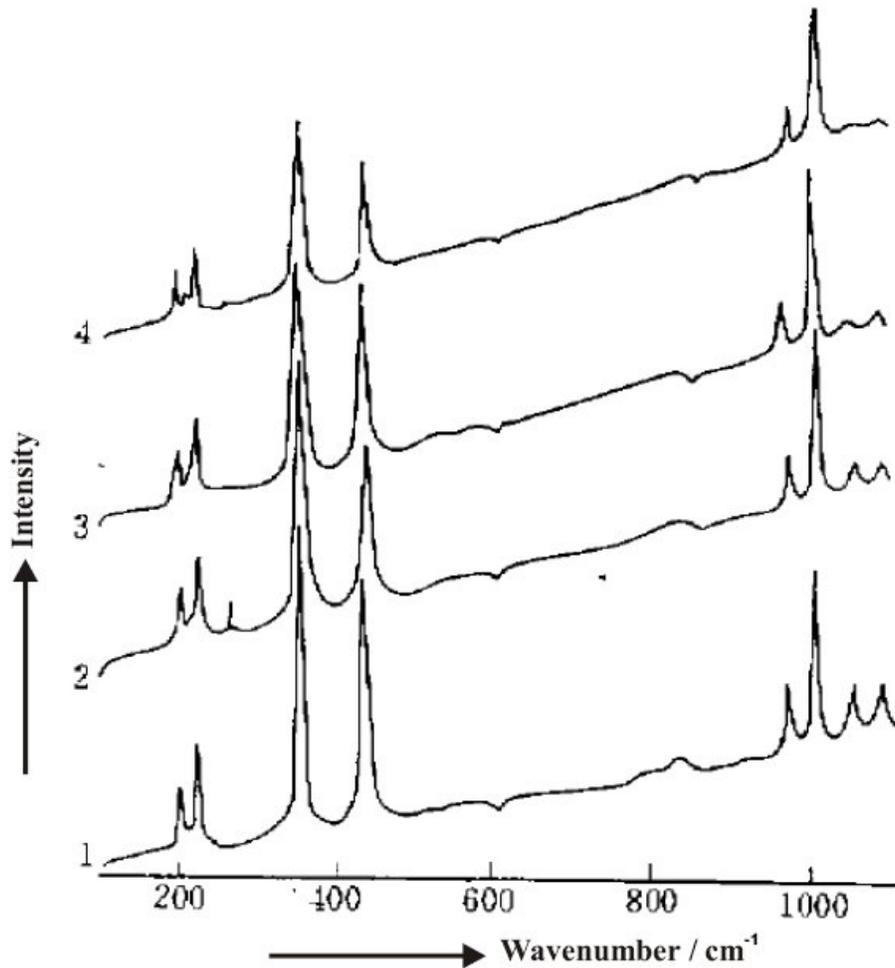

**Fig. 2. Raman spectra from core to rim (1→4) of M-y group of zircon crystal.**

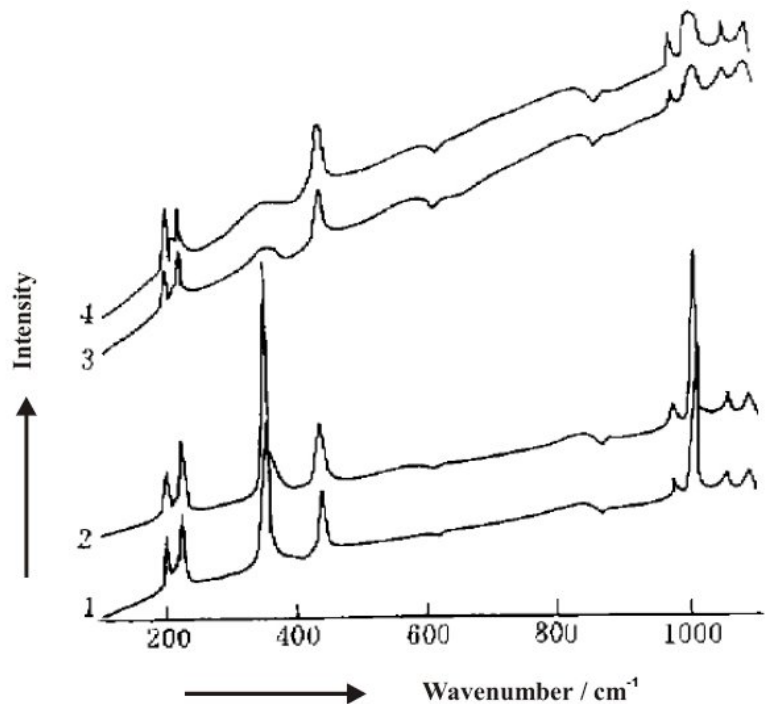

**Fig. 3 Raman spectra from core to rim (1→4) of magmatic zircon 9303 from pegmatite**

(2) magmatic zircons M-y and 9303 exhibit an obvious trend of decreasing Raman spectrum peak intensities from the core to rim of a crystal. On the other hand, metamorphic zircons M-b exhibit an obvious trend of increasing Raman spectrum peak intensities from core to rim of a crystal. Metamorphic zircons 87013 also exhibit a weak increase in Raman peaks 433cm[-1] and 937cm[-1]. This kind of reverse variation trend between magmatic and metamorphic zircons indicates that they have crystallized in opposite physical and chemical evolution conditions. In addition, metamorphic zircon 87013 in Raman spectrum peaks 368cm[-1] (A) and 439cm[-1](B) exhibit a trend of A > B → A = B → A < B from core to rim of a crystal. However, metamorphic zircons M-b exhibit a trend of A< B → A = B → A > B from core to rim of a crystal. This indicates that there may be a gradual structure and/or composition variation from core to rim of a crystal, which may reflect a gradual variation in their crystallization conditions.

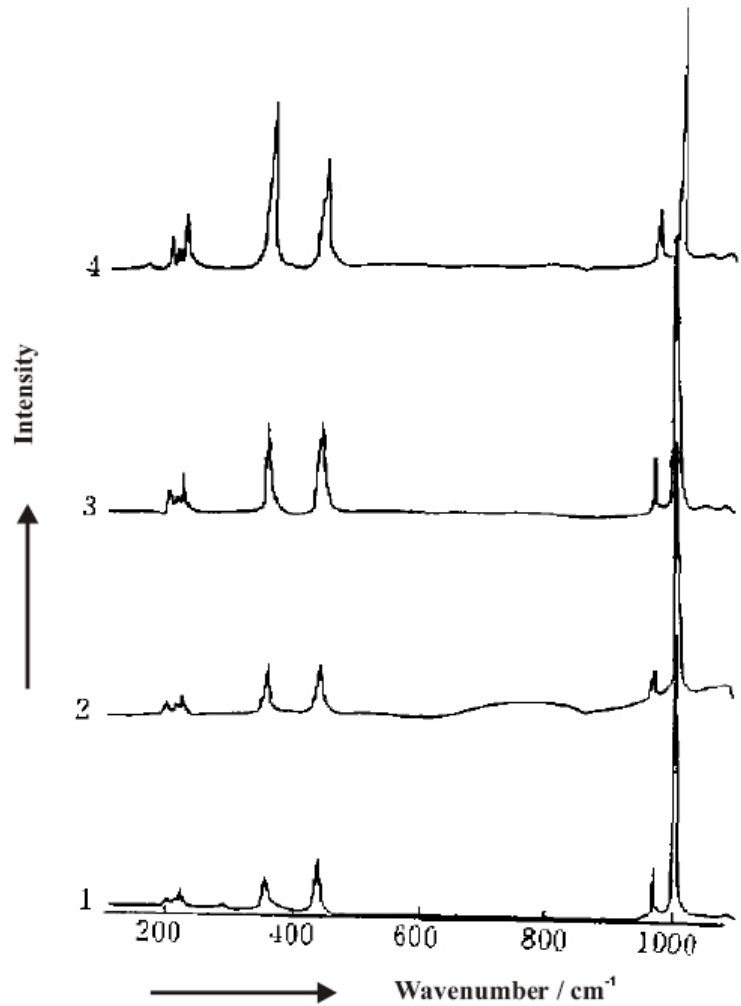

**Fig. 4. Raman spectra from core to rim (1→4) of a zircon crystal from M-b group.**

(3) As shown in table 2, metamorphic zircons exhibit a shift to the right (higher wavenumber side) in each Raman peak, especially the peaks Eg (III), $b_{2g}$ (II), $B_{1g}$ ($v_3$). On the contrary, magmatic zircons 9303 exhibit a shift to the left (lower wavenumber side) in each Raman peak. Magmatic zircons M-y1 and M-y2 exhibit none or no obvious shift in Raman peaks. We can not determine the factor(s) controlling these kinds of shifts.

## 3. CONCLUSIONS

(1) The zircons with an age of 2300 Ma± are of magmatic origin. Their age is the formation time of the parent rock. The zircons with an age of 400 Ma± are of metamorphic origin. Their age is the time when the parent rock experienced a metamorphic event.

(2) The magmatic zircons exhibit a trend of decreasing $ZrO_2/HfO_2$ ratio, and a trend of increasing $HfO_2$ concentration from the core to the rim of a crystal. Their Raman spectra have a high background level and exhibit a decrease of peak intensity from core to rim of a crystal. However, the medium to low grade metamorphic zircons exhibit a trend of increasing $ZrO_2/HfO_2$ ratio, and a trend of decreasing $HfO_2$ concentration from core to rim of a crystal. Their Raman spectra have a very low background level (the metamorphic zircons from higher metamorphic grade rocks – granulites, have relatively high background levels) and exhibit a decrease of peak intensity in all or part of the Raman spectrum peaks from core to rim of a crystal.

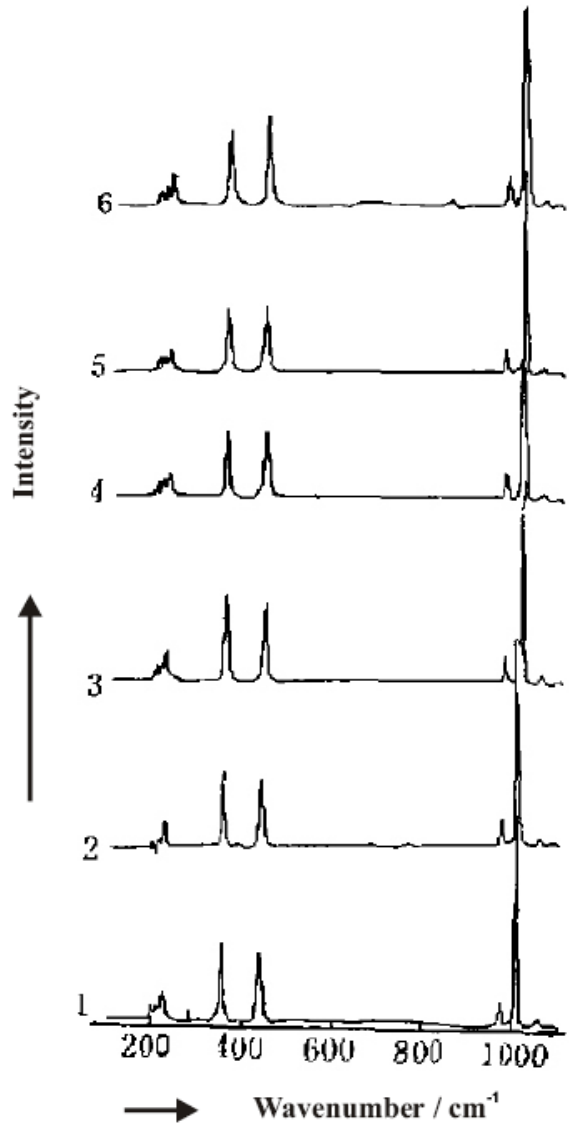

**Fig. 5. Raman spectra from core to rim (1 → 6) of metamorphic zircon 87013 from metamorphic granite.**

**Notes:**

This is our paper published originally in Acta Petrologica Et Mineralogica, 1996, Vol. 15, No. 1, 73-79.

These Raman spectrum zonations have been reexamined and the mechanism producing these zonations has been summarized in:

Bao et al., A Raman spectroscopic study of zircons on micro-scale and its significance in explaining the origin of zircons, Scientia Geologica Sinica, 1998, Vol.33 (4): 455-462, or at http://arxiv.org, arXiv: 0707.3184, July 2007.

In addition, a decreasing trend in crystallization temperature from core to rim of magmatic zircon crystals described in our papers is consistent with the calculated core-to-rim variation in crystallization temperature of a 4.3 billion year old magmatic zircon (Watson and Harrison, Science, 2005, V308, 841-844).

It is widely accepted that zircons can record the evolution history of the Earth and planets. For instance, through the studies of geochronology and isotope geochemistry of 4.0 - 4.4 billion years old zircons, it was derived that continental crusts began to form in Earth's first 100 million years (Ma) (Watson and Harrison, same as above; Harrison et al., Science, 2005, V310, 1947 – 1950), although they may have developed mainly around 2700 Ma and 1900 Ma ago (Condie et al., Precambrian Research, 2005, V139, 42-100).

The U, Th zonations of zircons described in our papers actually reflect the migration trends of U and Th in the Earth. U and Th are the most important heat-producing elements in the Earth and other terrestrial planets, and are the main energy sources for planetary evolution. At the same time, the X-rays released from them have an important influence on life beings. We have proposed the theoretical migration models and distribution patterns of U and Th in Earth and other terrestrial planets. Their influence on the Earth and terrestrial planetary dynamics, and the origin and evolution of life (including a graphic summary of zircon composition zonations and morphologies) has also been discussed in:

Bao X., Zhang, A., Geochemistry of U and Th and its influence on the origin and evolution of the Earth's crust and the biological evolution. Acta Petrologica et Mineralogica, 17(2): 160-172 (1998), or at http://arxiv.org, arXiv: 0706.1089, June 2007.

## Acknowledgements

I would like to thank Dr. RA Secco for offering me a postdoctoral position, which provoked my interest in translating this paper in my after-work time. The original work was supported by a grant awarded by the National Natural Science Foundation of China (49202021) to Xuezhao Bao.

## References

1. Gan X., 1995. The fluid inclusions in the metamorphic rocks in the Mayuan group: the P-T-t path during the rising process of rock. Contributions to Geology and Mineral Resources Research, (1): 38-47.

2. Black LP, Williams IS, Compston W, 1986. Four zircon ages from one rock: the history of a 3930 Ma-old granulite from Mount Sones, Enderby Land, Antarctica. Contributions to Mineralogy and Petrology, (94): 427- 437.

3. Wang X., Wang D., Zhou X., 1992. The morphology and its application of the re-crystallization zircons from Danzhu gneissoid granite.  Chinese Science Bulletin, (20): 1976-1879.

4. Bao X., 1995. Two kinds of composition variation treads of zircons and their ignificance in origin interpretation. Acta Mineralogica Sinica. 15(4): 404-410, or at http://arxiv.org, arXiv: 0707.3180, July 2007.

5. Bao X., Lu S., Li H., Li H., 1995. Minerageny of zircons from high-grade metamorphic rocks in inner Mongolia and Hebei. Acta Petrologica Et Mineralogica, 14( 3): 252-261.

6. Lin C., Tian S., 1994. Vibration spectra of zircons and the normal coordinate analysis. Acta Mineralogica Sinica. (2): 134-141.

7. Andersen, ME, Muggli, RZ. 1981. Microscopical techniques in the use of the molecular optics laser examiner Raman microprobe. Analytical Chemistry, 53: 1772-1777.

8. Mineralogy group, Wuhan college of geological sciences, 1979. Crystallography and mineralogy. Beijing: Geological publishing house, 187-188.

9. Gao S., 1987. The zircon features from two kinds of mineralized granites in Yanshan age. Acta Petrologica Sinica, (4): 63-71.